\documentclass{tTRA2e}

\usepackage{subfigure}
\usepackage{multirow}

\begin{document}



\title{Microscopic Travel Time Analysis of Bottleneck Experiments}

\author{M. Buk\' a\v cek$^{\rm a}$ $^{\ast}$\thanks{$^\ast$Corresponding author. Email: marek.bukacek@fjfi.cvut.cz \vspace{6pt}}, P. Hrab\' ak$^{\rm b}$ and  M. Krb\' alek$^{\rm a}$\\\vspace{6pt} $^{a}${\em{Czech Technical University in Prague, Faculty of Nuclear Sciences and Physical Engineering, Trojanova 13, Prague 2, 120 00 Czech Republic}};
\\ $^{b}${\em{Brno University of Technology, Faculty of Civil Engineering, Veve\v r\'i 331/95, 602 00 Brno, Czech Republic}}
\\\received{v1.0 released May 2016} 
}

\maketitle

\begin{abstract}

This contribution provides a microscopic experimental study of pedestrian motion in front of the bottleneck. Identification of individual pedestrians in conducted experiments enables to explain the high variance of travel time by heterogeneity of the crowd. Some pedestrians are able to push effectively through the crowd, some get trapped in the crowd for significantly longer time. This ability to push through the crowd is associated with the slope of individual linear model of the dependency of the travel time on the number of pedestrians in front of the bottleneck. Further detailed study of the origin of such ability is carried out by means of the route choice, i.e. strategy whether to bypass the crowd or to walk directly through it. The study has revealed that the ability to push through the crowd is a combination of aggressiveness in conflicts and willingness to overtake the crowd.

\begin{keywords} 
pedestrian dynamics; egress experiments; path analysis; travel time; aggressiveness; classification according to strategy
\end{keywords}

\end{abstract}

\section{Introduction}

In the last decades many  experiments were perfomed to describe, understand, and correctly model important aspects of pedestrian flow and crowd dynamics. Due to  variability of the problems investigated caused  mainly by the non-linear nature of pedestrian trajectories, the number of the phenomena described  is quite high, and often with uncertain quantification. An extensive review of the experiments and the phenomena observed and its use in model calibration can be found in \citep{HelBuzJohWer2005TS, SchKliKluKreRogSey2009, SchChoNis2010}

Particular focus is put on bottleneck flow, i.e., flow through an exit door, narrow corridor, and the like. A very important observation is that the flow continuously depends on the bottleneck width. Furthermore, within a certain reasonable range of width $w$, the dependence is considered to be linear. In  \citep{SeyPasSteBolRupKli2009TS} it is suggested that the dependence is $J=1.9\, w$. However, the actual slope of the dependence is highly influenced by crowd composition, motivation of people to pass, and, very notably, on the bottleneck type \citep{ZhaSey2014Procedia}.

The most straightforward way how to measure pedestrian flow is to calculate the number of passing pedestrians $\Delta N$ during a certain time $\Delta T$, and expressing this flow as ratio $J=\Delta N/\Delta T$. Such approach assumes the system to be in steady state, which is usually not trivial to detect \citep{LiaTorSeyChrDrzZheZha2016PhysicaA}. Moreover, to fully understand the crowd dynamics, it is necessary to express the actual flow $J(t)$ so as to detect the temporal evolution of the flow and smooth out simultaneously the measurement-induced fluctuation \citep{SteSey2010PhysA}.

Microscopic analysis of flow properties is usually performed under the assumption that pedestrians are indistinguishable, and, therefore,  the uncertainty of pedestrian reaction is modelled by statistical distribution of some key quantity as e.g. temporal headway $\Delta t$ between two consecutive pedestrians. Temporal headway distribution seems to be a very important characteristics of pedestrian flow closely related to the flow, since $J=\langle\Delta t\rangle^{-1}$. An extensive empirical study of temporal headway distribution for various bottleneck widths was undertaken by \citep{KreGruSch2006JSM}. Flow disturbance and distribution of delays caused by the clogging was studied by means of the social-force model in \citep{ParDor2005PhysA}. Recently in \citep{BodCod2016TGF} the authors have been investigating statistical models of temporal headway distribution with respect to the conditions in front of the bottleneck (density, angle of the nearest pedestrian, distance of the nearest pedestrian, etc.). Such approach tries to explain flow fluctuations by means of the variability of crowd composition in front of the bottleneck.

Crowd behaviour upstream the bottleneck was investigated by \citep{DuiDaaHoo2014Procedia} and \citep{BukHraKrb2015TGF}. The former investigated the pedestrian's anticipation, the latter the pedestrian's reaction to increasing density. Both papers are based on individual trajectories through the fundamental (velocity-density) diagram.

The ability to track individual trajectories in front of the bottleneck is crucial for micro-structure analysis of the pedestrian crowd. Experimental conditions and advanced image processing techniques enable to track not only the individual trajectories of pedestrians, but even to assign the trajectories to specific participants of the experiment, e.g., using individual markers \citep{MehBolSey2016TGF, MehBolMatLei2015LNCS}. In \citep{BukHraKrb2014Procedia} such approach was used to study the behaviour of individuals participating in the experiment. The experiment was repeated multiple times  and therefore provided records of individual participants in various situations. Thus it is possible to classify pedestrians with respect to personal preferences, abilities, or strategies. Their classification is based usually on the exit-choice strategy, as e.g. in \citep{HagSar2016TRR}.

Classification of pedestrians based on their personal properties is closely related to the term of crowd heterogeneity, which can be studied from a variety of aspects: velocity, ability to push through the crowd, and preferred route choice. Heterogeneity of pedestrians significantly affects the quality of flow \citep{CamHooDaa2009TRR}, and ``diversity of pedestrians is crucial for phase separation'' \citep{SeyPorSch2010LNCS}. In~\citep{BukHraKrb2014Procedia}, heterogeneity was studied by means of the  dependence of the travel time (time spent in the room) on the size of the crowd in front of the bottleneck . Here the slope of such dependence reflects the ability of individual pedestrians to push through the crowd (referred to as aggressiveness). However, in \citep{BukHraKrb2016TGF} we discovered that such ability is not necessarily related to a pedestrian's aggressiveness but also to the preferred route choice, i.e. the decision to push through the crowd directly or bypassing it (i.e. walking around the crowd and reaching the exit by squeezing between the crowd and the wall). The main goal of this paper is to extend the previous studies by a classification of pedestrians with respect to the strategies  observed. The key quantity is the time spent in the room and its dependence on the above mentioned aspects and individual's properties.

\section{Preliminaries}

\subsection{Experiments}

This paper investigates data from two egress experiments organized at the Czech Technical University in Prague. Specifically, the study is mainly based on the “passing-through” experiment conducted on 29th April 2014 (for consistency with other publications referred to as E4) for the purpose of boundary-induced phase transition study, see \citep{BukHraKrb2014Procedia}. The individual markers used enabled  to carry out extraction of trajectories automatically. Some results are also supported by data from the  earlier ``passing-through'' experiment conducted on 10th December 2012 (refered to as E2), see  \citep{BukHraKrb2015TGF}. Nevertheless, only a short video-footage from this experiment was processed manually for this study, and therefore it gives only aggregated information without individual aspects of pedestrians (there are at most 6 paths for each participant, which is not sufficient for a deeper analysis).

\begin{figure}[h!]
\begin{center}
\subfigure[Set up of Experiment]{
\resizebox*{9cm}{!}{
	\includegraphics[height=9cm]{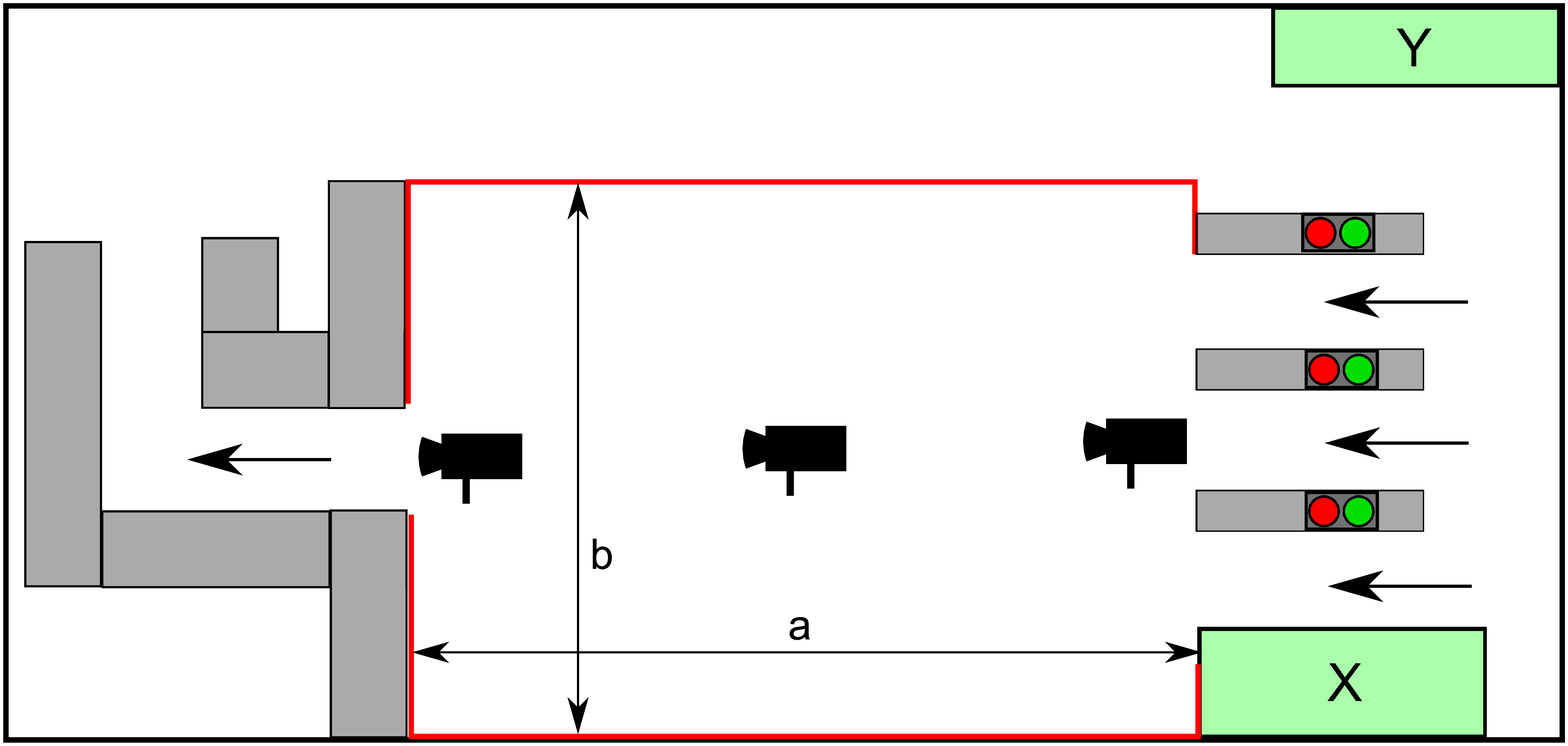} }}\hspace{5pt}
\subfigure[Snapshot of Experiment]{
\resizebox*{4.5cm}{!}{
	\includegraphics[height=4cm]{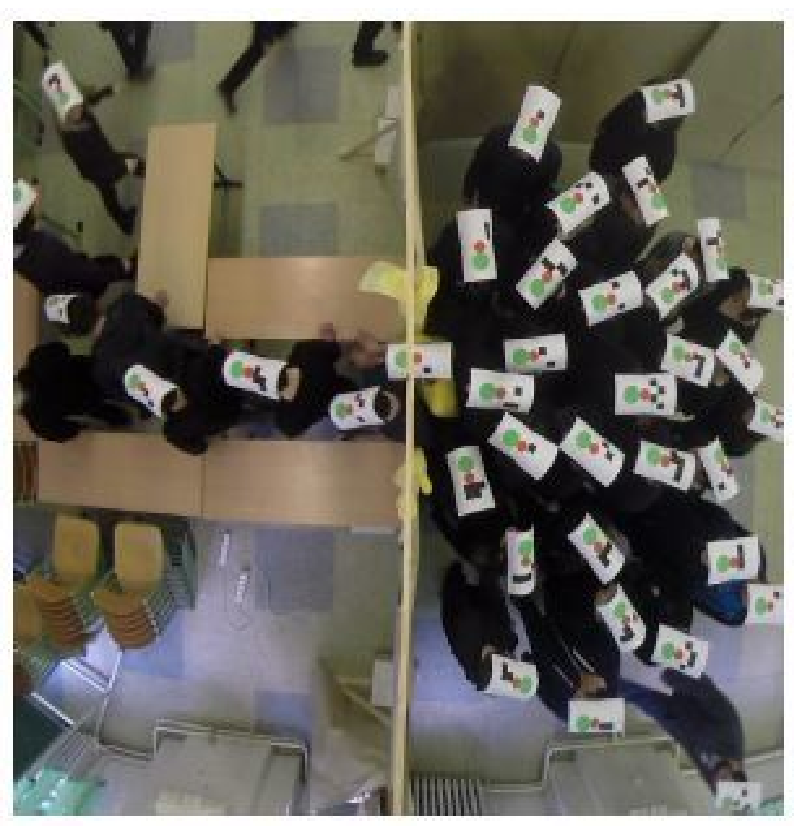}
	}}
	\caption{Left: schematic representation of the experiment. The entrance-exit distance $a = 7.2$~m (considered for measuring travel time), room width $4.5$~m, and exit width $0.6$~m. Right: snapshot from exit camera.}
\label{fig:ExpIlust}
\end{center}
\end{figure}

Let us describe the idea of experiment E4. A group of 76 volunteer students was passing repetitively through an artificial room, whose layout is illustrated in Fig.~\ref{fig:ExpIlust}. The design of the experiment was as follows. The volunteers were instructed to enter the room, pass through it as fast as possible avoiding running, and then return to the pedestrian cluster in front of the entrance. This technique enabled to maintain constant flow through the room. The inflow rate was controlled using three independent signalling devices informing pedestrians in the crowd to enter the room trough one out of three available entrances. To simulate random inflow conditions, green light was alternated by $k\cdot \Delta h$ seconds of red light, where $k$ was generated from geometric distribution; $\Delta h = 0.6$~s was the minimal time step, to which pedestrians were able to react reliably. Each run of the experiment started with an empty room.

The pedestrians were wearing contrast hats enabling automatic extraction of individual trajectories. Also, there were unique codes on the hats enabling identification of the pedestrians; therefore every trajectory was assigned to some participant. This arrangement was very useful for comparison of individual participants and of their behaviour.

\subsection{Definitions}

This article works with two main objects of investigation: paths and participants. By the term path we understand one recorded passing of some individual through the room. The path will be denoted by lower-case Latin letters, mainly $i$, and the index set of all paths will be denoted by upper-case letter $I$. Each path $i\in I$ is assigned a trajectory $\vec{x}_i$ consisting of time-space coordinates, i.e.,
\begin{equation}
	\vec{x}_i = \{(x_i(t),y_i(t),t)\mid t\in T_i\},
\end{equation}
where $x_i(t)$ and $y_i(t)$ are coordinates of paths $i$ in time $t$. The time set $T_i=\big[\,T_\mathrm{in}(i),T_\mathrm{out}(i)\,\big]$ covers the period from entrance to the exit of one passing, i.e.,$T_\mathrm{in}(i)$ is the time of the beginning of path $i$ measured at the entrance and $T_\mathrm{out}(i)$ is the time of the end of $i$ measured at the exit. Here we note that the path identifier $i$ does not refer to any specific participant, but only to the recorded path.

The term participant refers to a specific participant of the experiment. Semi-automatic detection enabled us to distinguish individual participants of the experiment and extract individual characteristics of participants, because there are several trajectories assigned to each participant. Participants are denoted by Greek letters, mainly $\alpha$, and the index set of all participants is denoted by $A$. The index set of all trajectories associated with participant $\alpha\in A$ is denoted $I_{\alpha}\subset{I}$.

The main quantity studied is \textbf{Travel time} defined as the time spent in the room. The travel time $TT(i)$ assigned to path $i\in I$ is defined as
\begin{equation}
	TT(i)=T_\mathrm{out}(i)-T_\mathrm{in}(i)\,.	
\end{equation}

Another quantity influencing the path's properties is the mean occupation of the room denoting the average number of pedestrians in the room. The mean occupancy $\overline{N}{(i)}$ assigned to path $i$ is defined as time average of the actual occupancy $N(t)$, i.e.,
\begin{equation}
	\overline{N}(i)=\frac{1}{TT(i)}\int_{T_\mathrm{in}{(i)}}^{T_\mathrm{out}{(i)}}N(t)\mathrm{d}t\,.
\end{equation}
Here we note that the integral in the definition above is rather formal, because the camera records are limited by the frame-rate, and, technically, the average occupancy is calculated using a sum. On the other hand, this representation enables to calculate the mean occupancy even for event driven   updates or records. For readability reasons, the mean occupancy $\overline{N}$ in the graphs is denoted as \verb+N_mean+.

To compare the travel time measured under various conditions (although the inflow parameters were fixed for the whole run, the conditions inside the room were changeable), scaling based on mean occupancy is introduced. For each occupancy bin $(N-1,N], \, N \in 1 \ldots N_{\mathrm{max}}$,  the mean travel time $TT_N$ is defined as
\begin{equation}
	TT_N = \operatorname*{average} \left\{TT(i) \mid i\in I,\,\, \overline{N}(i)\in (N-1,N]\right\}\,.
\end{equation}
Then the relative travel time $TT_R(i)$ for each path $i\in I$ is defined as
\begin{equation}
	TT_R(i) = \frac{TT(i)}{TT_N}\,.
\end{equation}

\begin{figure}[h!]
\begin{center}
	\includegraphics[scale=.6]{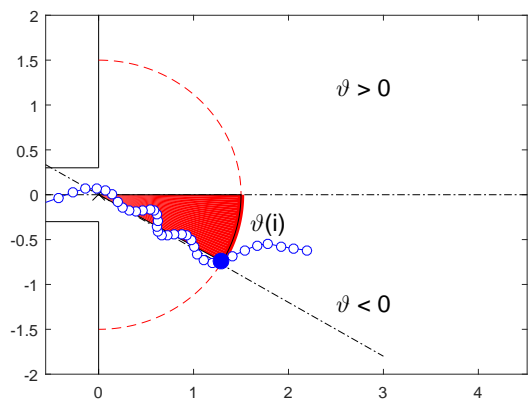}
	\caption{Illustration of exit angle definition.}
\label{fig:angleDef}
\end{center}
\end{figure}

As shown below, an important property of path $i\in I$ is deviation of the trajectory from straight direction towards the exit at which the pedestrian joined the crowd in front of the exit. Since the moment of joining the crowd is automatically hard to recognize, the angle used for the analysis was the angle while the pedestrian was passing a semicircle with center in the exit $\vec{e}$ and radius of $1.5$~m. The measurement of exit angle $\vartheta(i)$ is illustrated in Fig.~\ref{fig:angleDef}. The angle of direct trajectory from the entrance to the exit is $\vartheta \sim 0 \deg$, angles of trajectories along the walls are $\vartheta \sim \pm 80 \deg$; the positive sign refers to the right-hand side with respect to the flow direction.

\subsection{Basic Observations}

The overview of all runs considered in our experiments is given in Tab.~\ref{tab:makro}. Inflow in the experiments was controlled in order to obtain information about the behaviour of pedestrian crowd under a  variety of conditions,and thus we have at our disposal data from a variety of regimes (or phases).

\begin{table}[h!]
	\caption{Summary of runs performed. $J_{\rm in}$ and $J_{\rm out}$ refer to flow measured at the entrance and at the exit, respectively; $\overline{TT}$ is mean travel time in a given run. $N(150)$ specifies the number of pedestrians in the room 150 s after initiation; {\#}paths denotes the number of passings in a given run. The values of run 7 of E2 are only tentative.}
	\label{tab:makro}
\begin{tabular}{rcccccl}
	run	& $J_{\rm in}$ [ped/s] & $J_{\rm out}$ [ped/s] & $\overline{TT}$ [s]& $N(150)$ [ped] & {\#}paths & Observation\\
	\hline\hline	
	\#~~2 &0.99 &0.99 &\hspace{1.1ex}5.67 &\hspace{1.1ex}3 &158 &free flow\\
	\#~~5 &1.22 &1.20 &\hspace{1.1ex}6.73 &\hspace{1.1ex}7 &274 &free flow\\
	\#~~4 &1.37 &1.30 &16.59 & 24  &294 &stable cluster\\
	\#~~3 &1.43 &1.33 &14.39 & 22  &260 &stable cluster\\
	\#~~6 &1.39 &1.31 &20.40 & 33  &270 &stable cluster\\
	\#~~7 &1.55 &1.37 &25.78 & 45  &260 &transition\\
	\#~11 &1.61 &1.38 &21.65 & 41  &141 &transition\\
	\#~~9 &1.78 &1.37 &24.06 & 47* &148 &congestion\\
	\#~~8 &1.79 &1.38 &25.03 & 46* &144 &congestion\\
	\#~10 &1.78 &1.37 &23.33 & 44* &214 &congestion\\
	\hline
	E2~7  &1.50 & -   & 11.00    & 25   &452 &congestion\\
	\hline
\end{tabular}	
\end{table}

\begin{figure}[h!]	
\begin{center}
	{\includegraphics[scale=0.8]{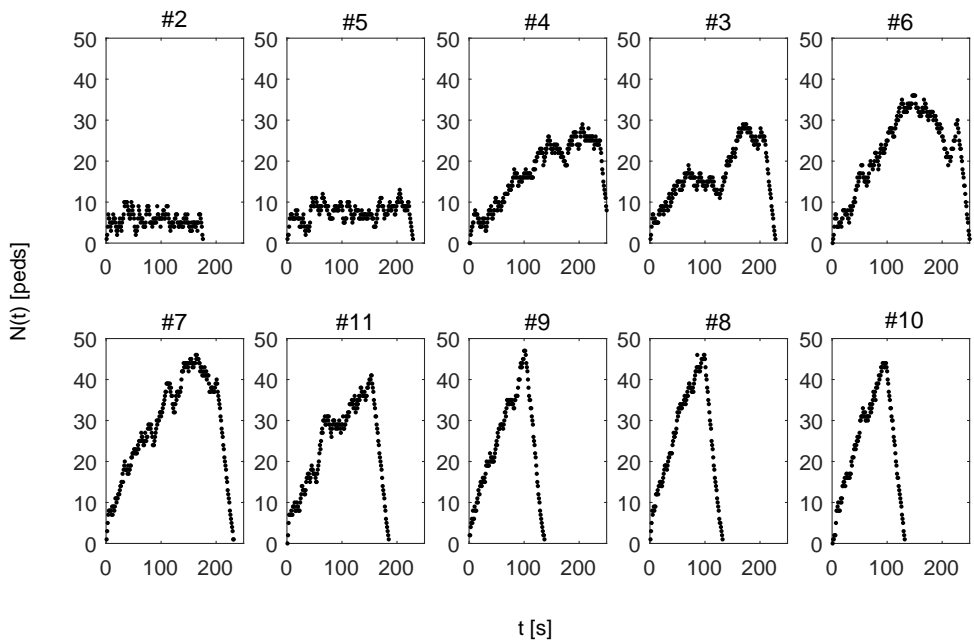}}
	\vspace{-1cm}
\caption{Temporal evolution of the number of pedestrians in the room for individual runs of experiment E4.}
\label{fig:ObsKola}
\end{center}
\end{figure}

Here it is important to note that the steady state has not been reached in every run.This is the reason why inflow $J_\mathrm{in}$ is slightly higher than outflow $J_\mathrm{out}$. The temporal evolution of the room occupancy is depicted in Fig.~\ref{fig:ObsKola}.

The term free-flow denotes a situation where interactions between pedestrians were rare, and there was no clogging in front of the exit. After increasing the inflow, the interaction started to play role, and small, but stable cluster has been formed in front of the exit. When the outflow reached  $1.38$~ped/s, the cluster started to grow and would fill the whole room, which would result in significant congestion.

In the previous papers, this experiment was used to describe the boundary-induced phase transition and its properties. To prove the existence of the transition, it was not only the evolution of occupancy that was used – another important quantity was the average travel time through the room, which is closely related to the flow.

\section{Travel-time Analysis}

This paper focuses on the properties of travel time $TT$. Despite previous studies, the data are not aggregated with respect to an individual run, but are classified irrespective of inflow/outflow conditions. Our aim is to show that travel time is mainly influenced by the size of clogging in front of the exit represented by mean occupancy $\overline{N}$.

\begin{figure}[h!]
\begin{center}
	{\includegraphics[scale=1]{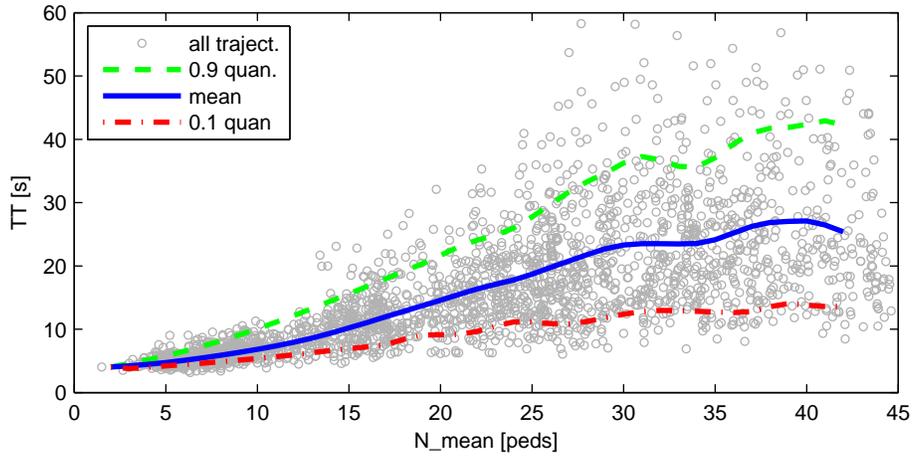}}
	\caption{Travel time -- occupancy relation, mean, top $10$~\% and bottom $10$~\% quantiles are highlighted (Experiment E4).}
\label{fig:TToccup}
\end{center}
\end{figure}

For each path $i\in I$, the pair $(\overline{N}(i),TT(i))$ is plotted in~Figure \ref{fig:TToccup}. Dividing the trajectories according to occupancy $\overline{N}$ into groups corresponding to $(N-1,N]$, we can analyse the dependency on the occupancy in more detail. The mean travel time $TT_N$ increases almost linearly with increasing occupancy $N$.

\begin{figure}[h!]
\begin{center}
\subfigure[Free flow]{
\resizebox*{6cm}{!}{\includegraphics[scale=1]{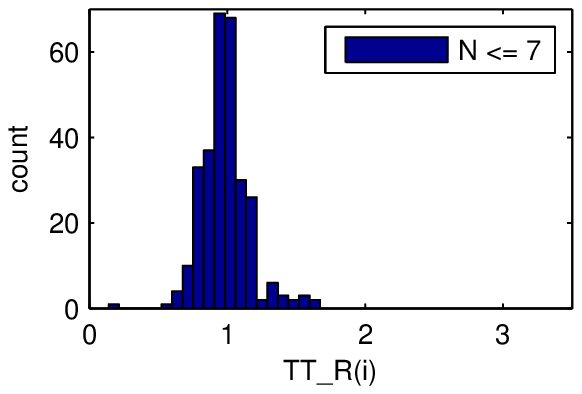}}}\hspace{5pt}
\subfigure[Congested]{
\resizebox*{6cm}{!}{\includegraphics[scale=1]{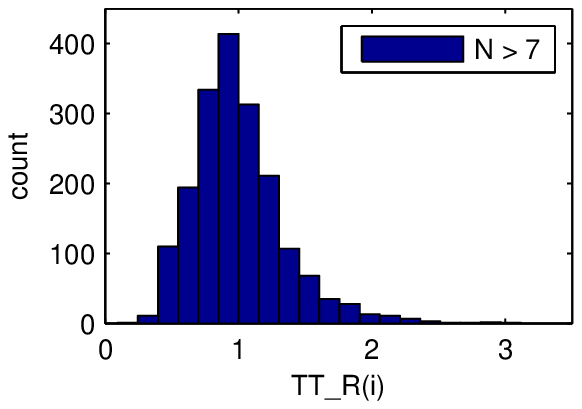}}}
\caption{\label{fig1} Histograms of relative travel time; data are filtered over different regimes. Experiment E4.}
\label{fig:histTT}
\end{center}
\end{figure}

This linear increase in mean travel time $\overline{TT}$ is accompanied by increasing variance of the measured travel-time, which is not proportional to the absolute value of the mean. Indeed, comparing the relative travel time of free flow, i.e. without interactions ($\overline{N}\leq7$),  with the relative travel time corresponding to the congestion regime, i.e. involving interactions ($\overline{N}>7$), we can observe a significant increase in variance, as shown in Fig.~\ref{fig:histTT}.

\begin{figure}[h!]
\begin{center}
	{\includegraphics[scale=1]{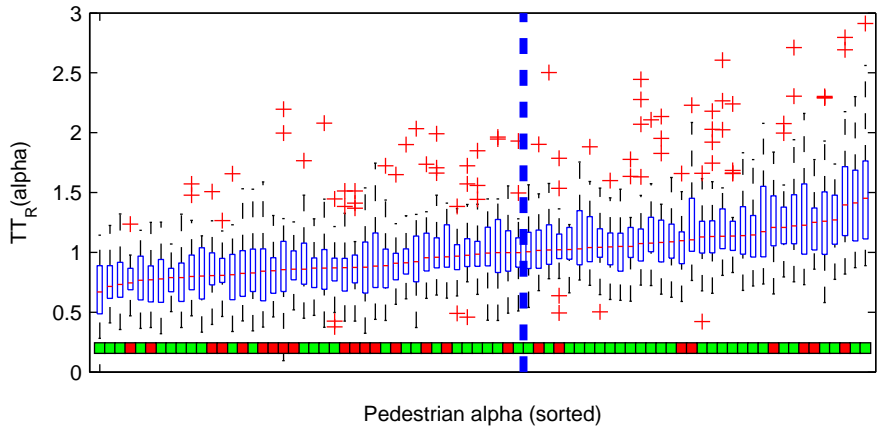}}
	\caption{Boxplot of $TT$ for each pedestrian, ordered by mean value. Coloured squares represent gender (red -- women, green -- men). Experiment E4.}
\label{fig:boxTT}
\end{center}
\end{figure}

This fact motivated us to investigate the dependence from the point of view of individual pedestrians $\alpha\in A$. Let us denote by $TT_{R,\alpha}$ the set of all relative travel times of paths corresponding to pedestrian $\alpha$, i.e. $TT_{R,\alpha}=\{TT_R(i)\mid i\in I_{\alpha}\}$. Box-plots of $TT_R$ related to all individuals are plotted in Fig.~\ref{fig:boxTT}; the participants are ordered according to average values of $TT_{R,\alpha}$.

From this graph it is evident that the increasing variance in $TT$ is caused by heterogeneity of individual properties of individual participants. This motivates us to handle the $TT-\overline{N}$ dependence separately for each participant.

\begin{figure}[h!]
\begin{center}
	{\includegraphics[scale=1]{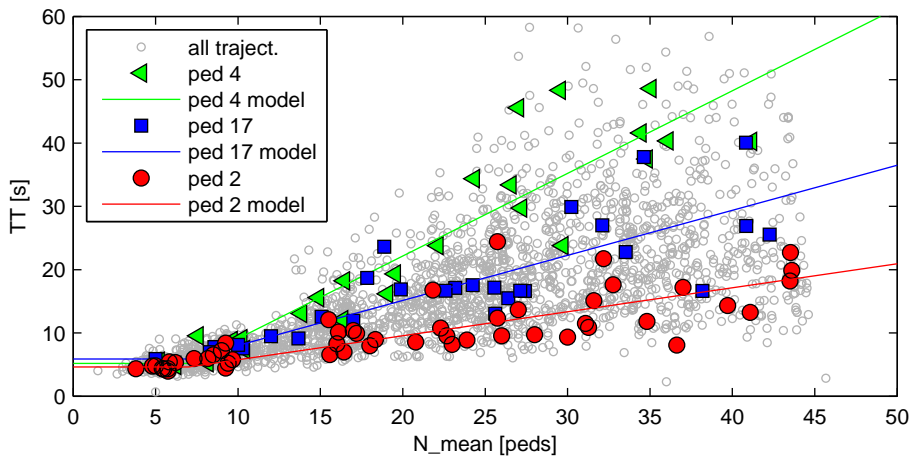}}
	\caption{Dependency of TT on occupancy, highlighting three pedestrians with evaluated linear model. Experiment E4.}
\label{fig:TToccupMod}
\end{center}
\end{figure}

Let us now consider one arbitrary but fixed pedestrian $\alpha$. We assume that the relation between $TT(i)$ and $\overline{N}(i)$ for $i\in I_\alpha$ can be expressed by means of a piece-wise linear model 
\begin{equation}
\label{eq:LMTT}
	TT(i)= \mathbf{a}_\alpha+ \mathbf{1}_{\{\overline{N}(i)>7\}}(\overline{N}(i)-7) \cdot \mathbf{b}_\alpha + \mathrm{noise}\,,
\end{equation}
where intercept $\mathbf{a}_\alpha$ can be understood as $\mathbf{a}_\alpha=S/v_{0,\alpha}$\,, $S$ being the entrance-exit distance, and $v_{0,\alpha}$ the desired free-flow velocity. Parameters $\mathbf{a}_\alpha$ and $\mathbf{b}_\alpha$ are unique for each participant. Factor $(\overline{N} - 7)$ has been derived from the data in order to distinguish situations without and with interactions. From the analysis it follows that for $\overline{N}\leq7$, the pedestrians are scattered in the room without significant interaction. If $N>7$, a cluster or crowd of pedestrians is formed in front of the exit. Values corresponding to three pedestrians can be seen in Fig.~\ref{fig:TToccupMod}.

The inverse of slope $\mathbf{b}_\alpha$ can be interpreted as the ability of pedestrian $\alpha$ to push trough the crowd in order to reach a lower travel time. The steeper the slope is, the more the values influence one another.

Correctness of the linear model has been checked using $R^2$ statistics, defined as $1-\mathrm{var}(TT-\mathrm{model})/\mathrm{var}TT$. The closer the statistics is to one, the more the variance in data is explainable by the linear model. The average $R^2$ statistics is $0.688$ with a minimum $0.386$ and maximum $0.936$. The low values of $R^2$ statistics were obtained for pedestrians with a low value of slope $\mathbf{b}_\alpha$, which is expected, because low slope means that the dependence of $TT$ on $\overline{N}$ is not significant.

\section{Route Choice}

Let us investigate more deeply the aspects explaining the variety of slopes of the linear dependence of $TT$ on $\overline{N}$. An important aspect influencing $TT$ is the route choice, more specifically, whether the participant pushes through the crowd directly or rather tries to bypass it, i.e. walks around the crowd and squeeze between the crowd and the wall.

\begin{figure}[h!]
\newlength{\iw}
\setlength{\iw}{3cm}
\begin{tabular}{cccc}
\hline
	fast in crowd & slow in crowd & fast in crowd & slow in crowd\\		
	$\overline{N}\in[25,35]$&$\overline{N}\in[25,35]$&$\overline{N}\in[35,50]$&$\overline{N}\in[35,50]$\\
	$TT<10~s$&$TT\geq35~s$&$TT<15~s$&$TT\geq35~s$\\
	\framebox{\includegraphics[width=\iw, angle = 180, clip,trim=8cm 5cm 7cm 3cm]{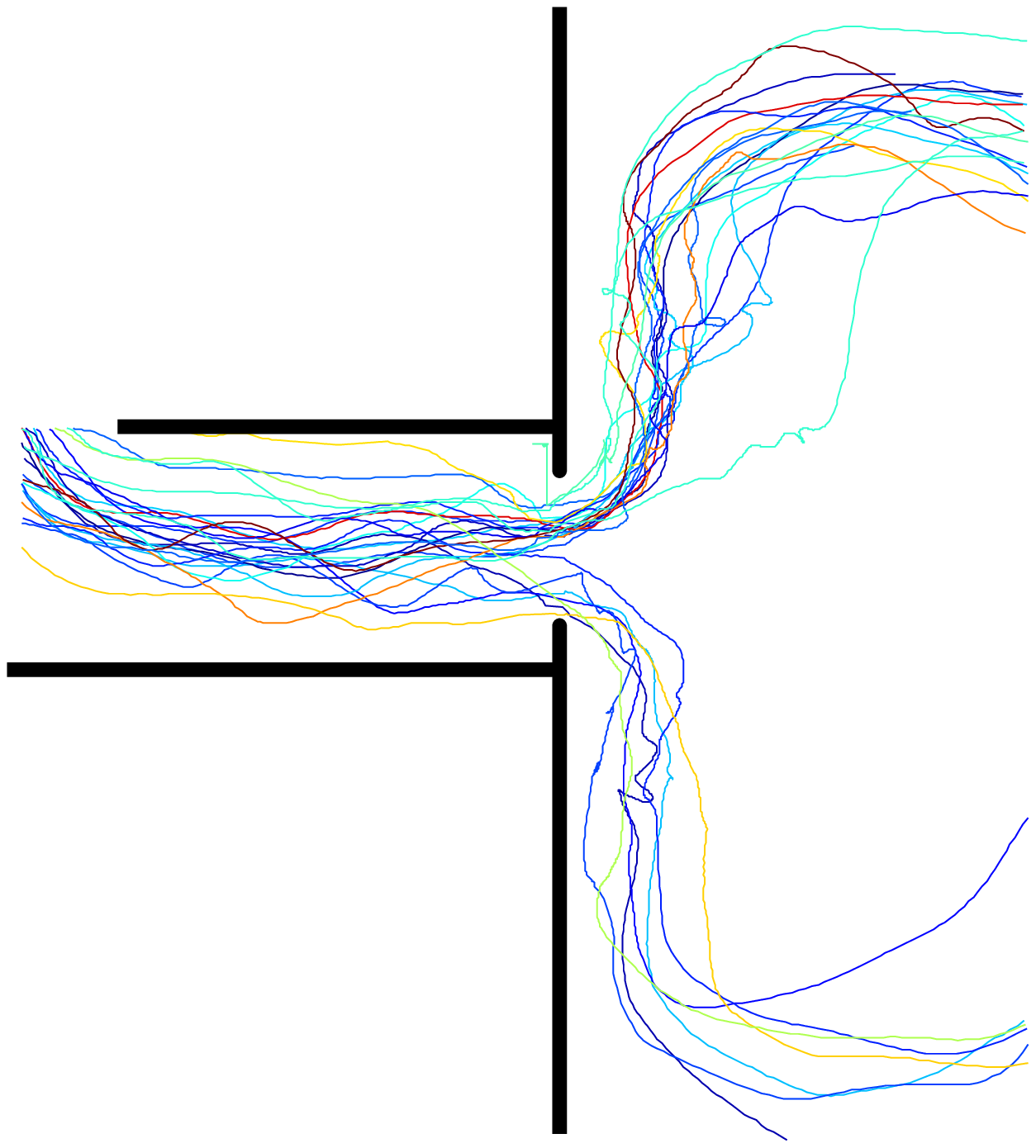}}&
	\framebox{\includegraphics[width=\iw, angle = 180, clip,trim=8cm 5cm 7cm 3cm]{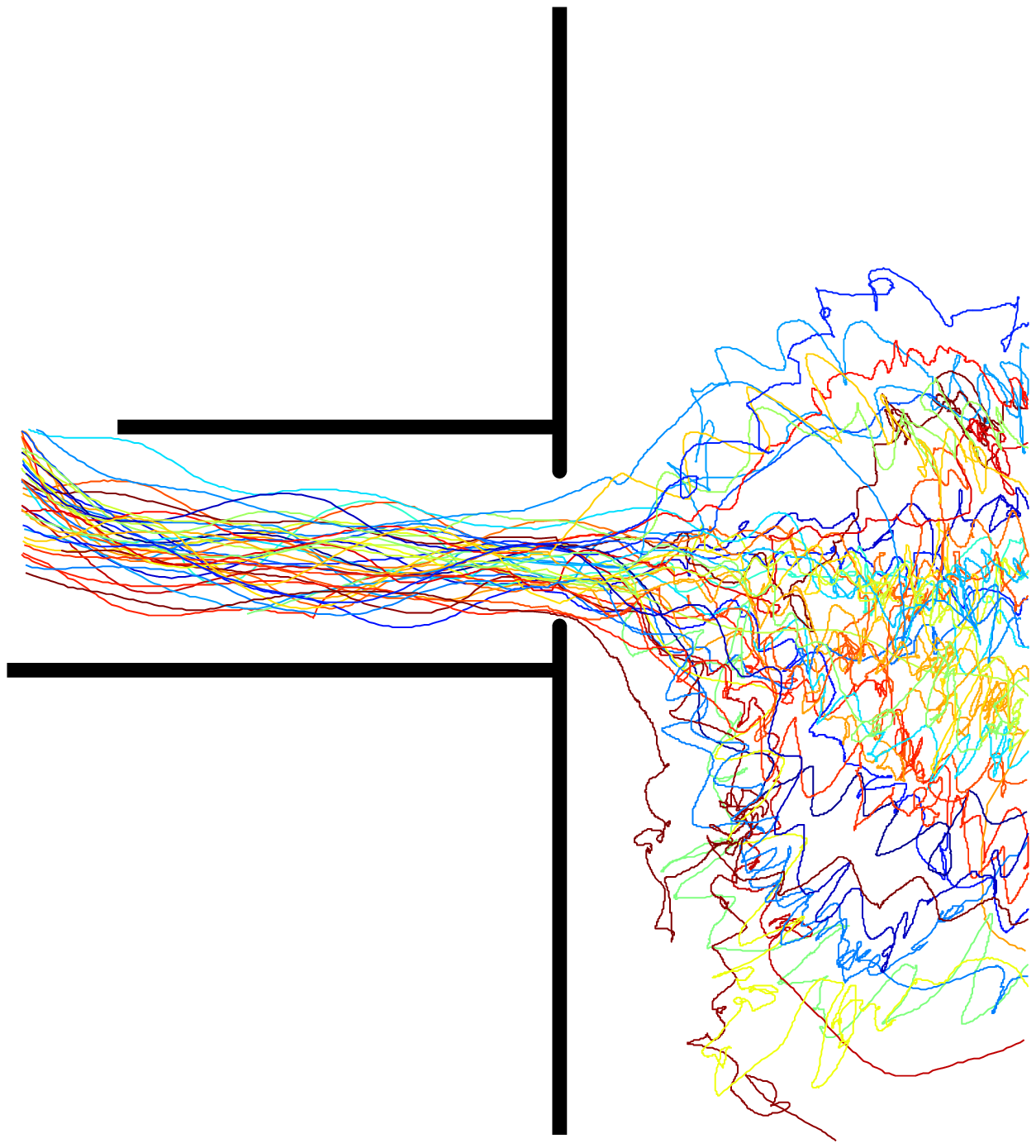}}&
	\framebox{\includegraphics[width=\iw, angle = 180, clip,trim=8cm 5cm 7cm 3cm]{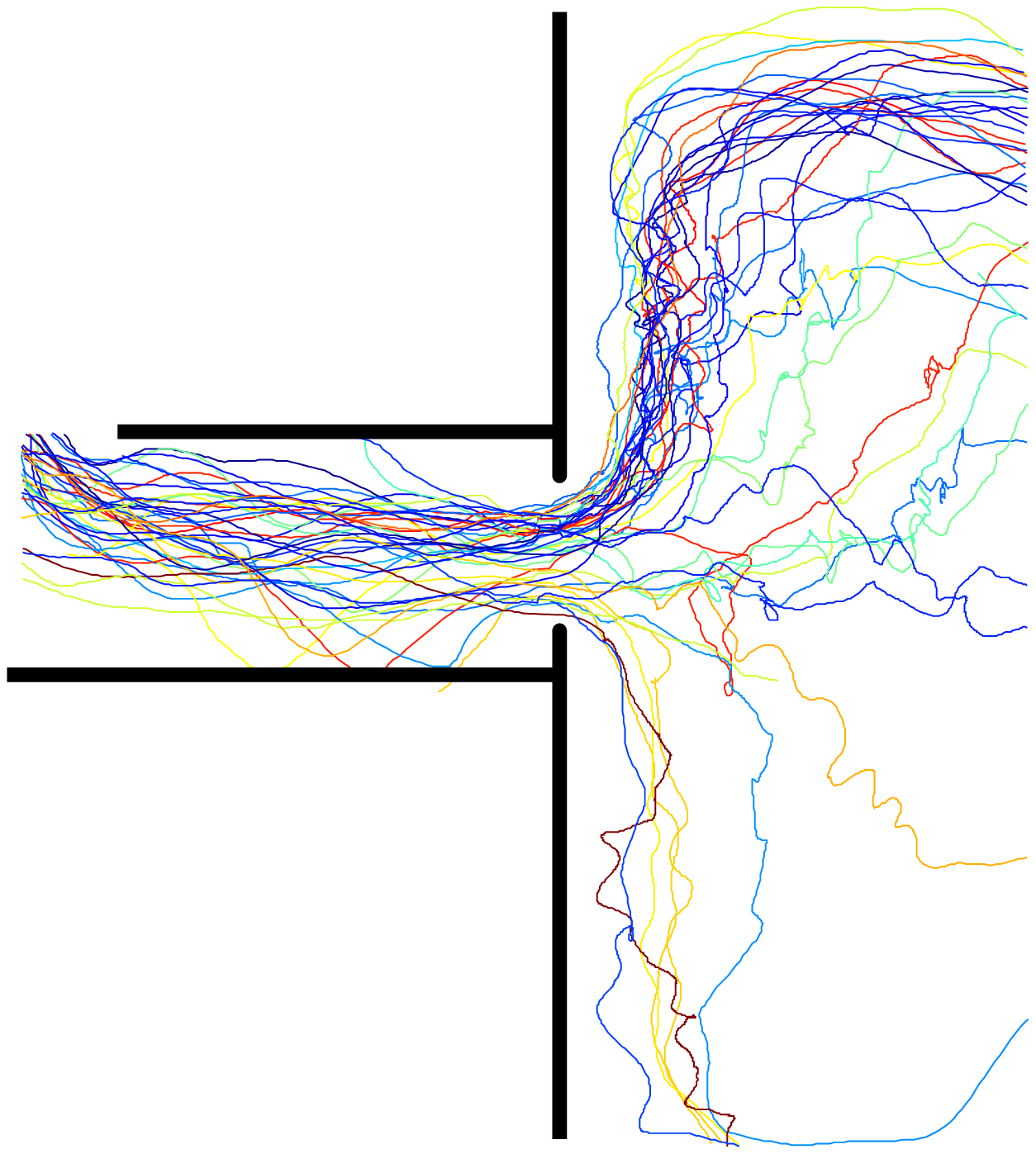}}&
	\framebox{\includegraphics[width=\iw, angle = 180, clip,trim=8cm 5cm 7cm 3cm]{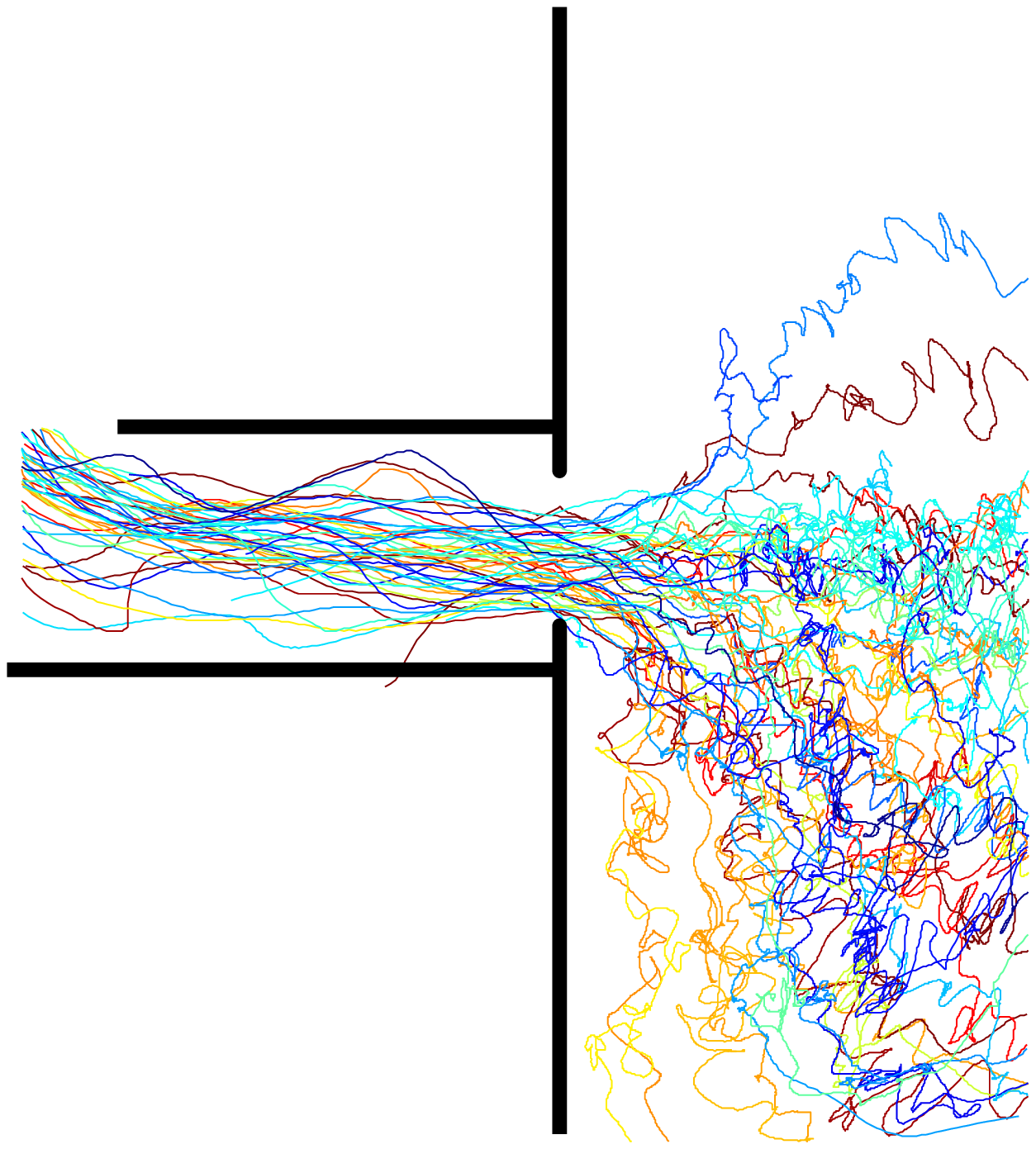}}\\
	
	\framebox{\includegraphics[width=\iw, angle = 180, clip,trim=8cm 5cm 7cm 3cm]{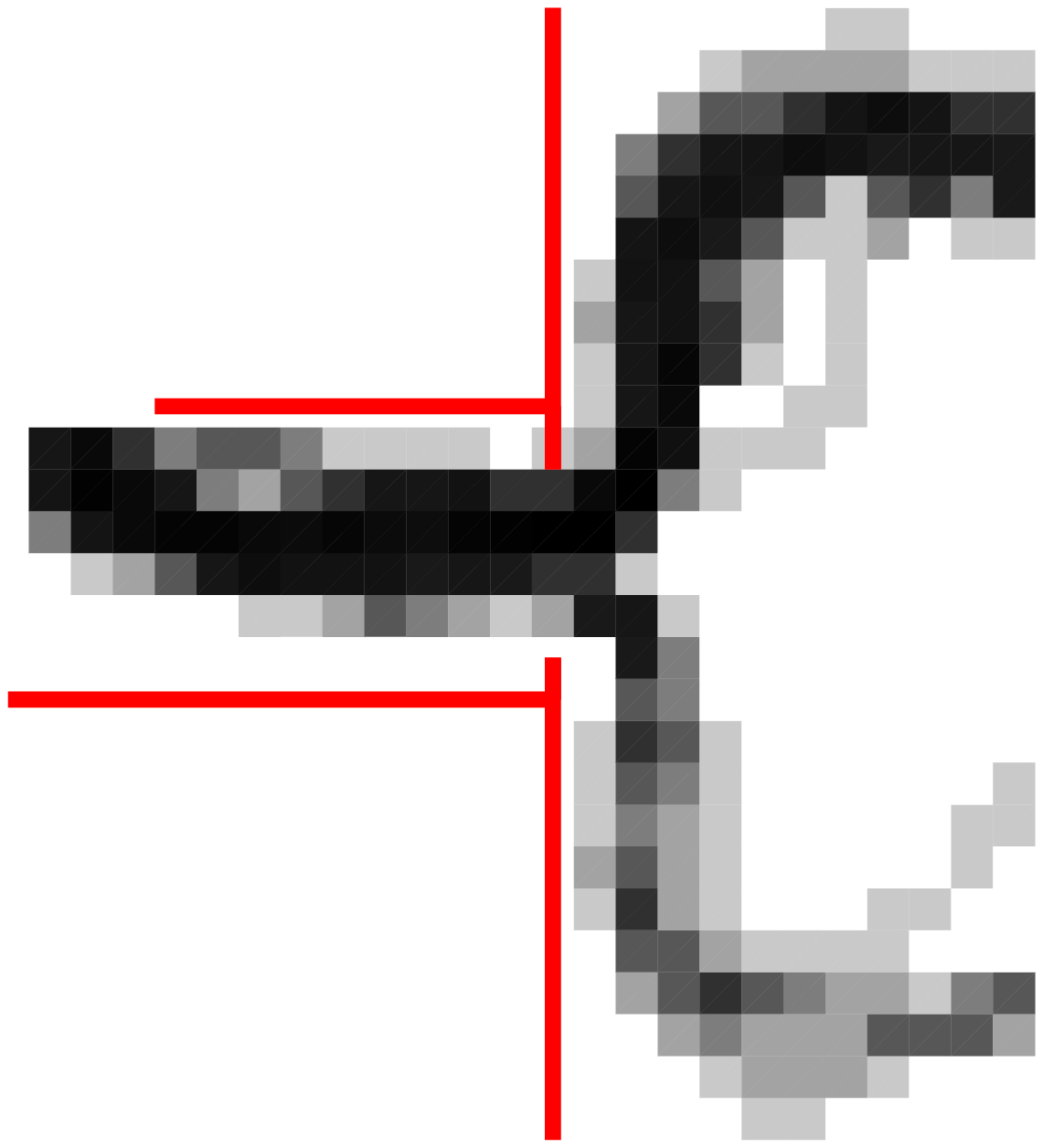}}&
	\framebox{\includegraphics[width=\iw, angle = 180, clip,trim=8cm 5cm 7cm 3cm]{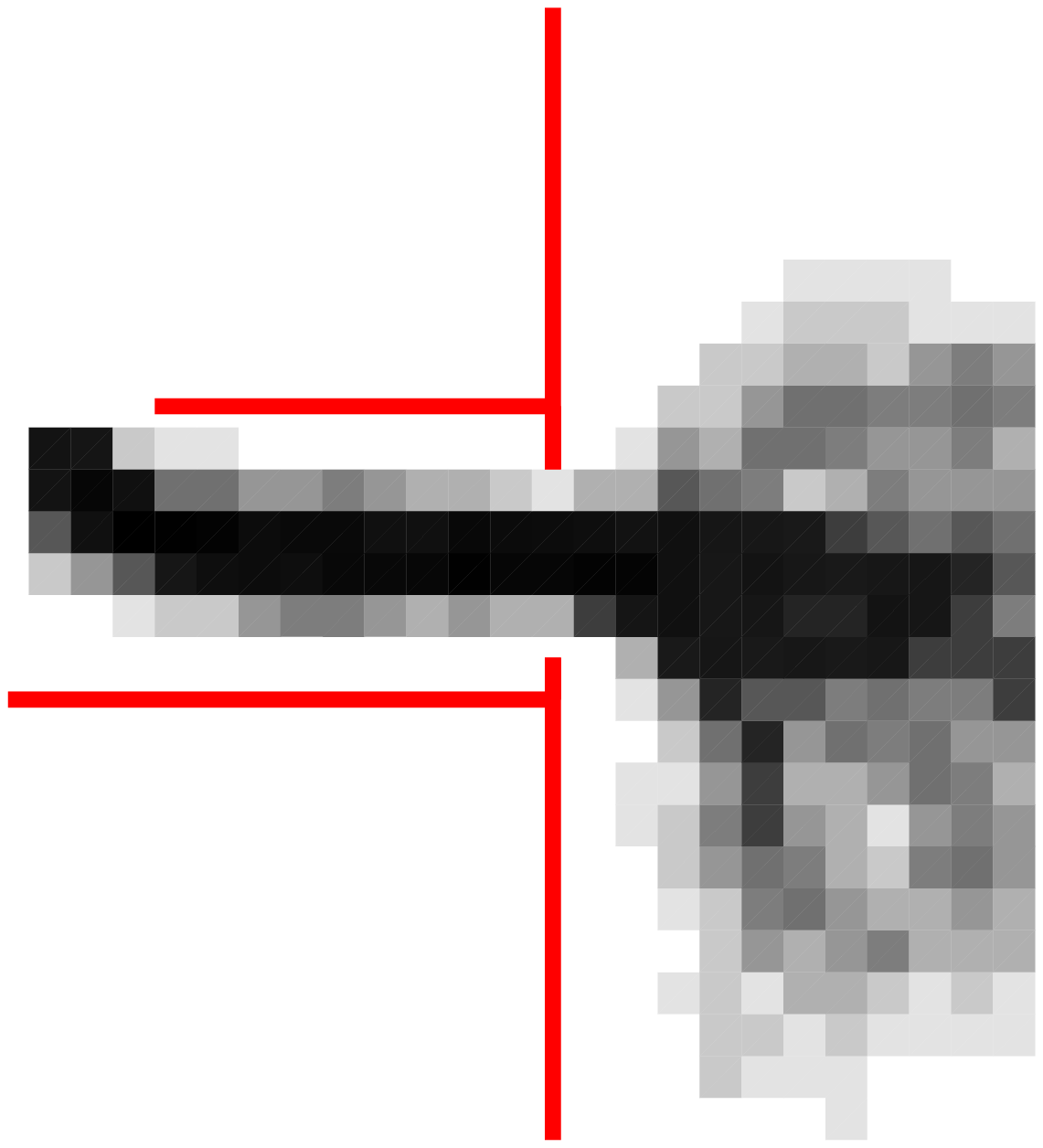}}&
	\framebox{\includegraphics[width=\iw, angle = 180, clip,trim=8cm 5cm 7cm 3cm]{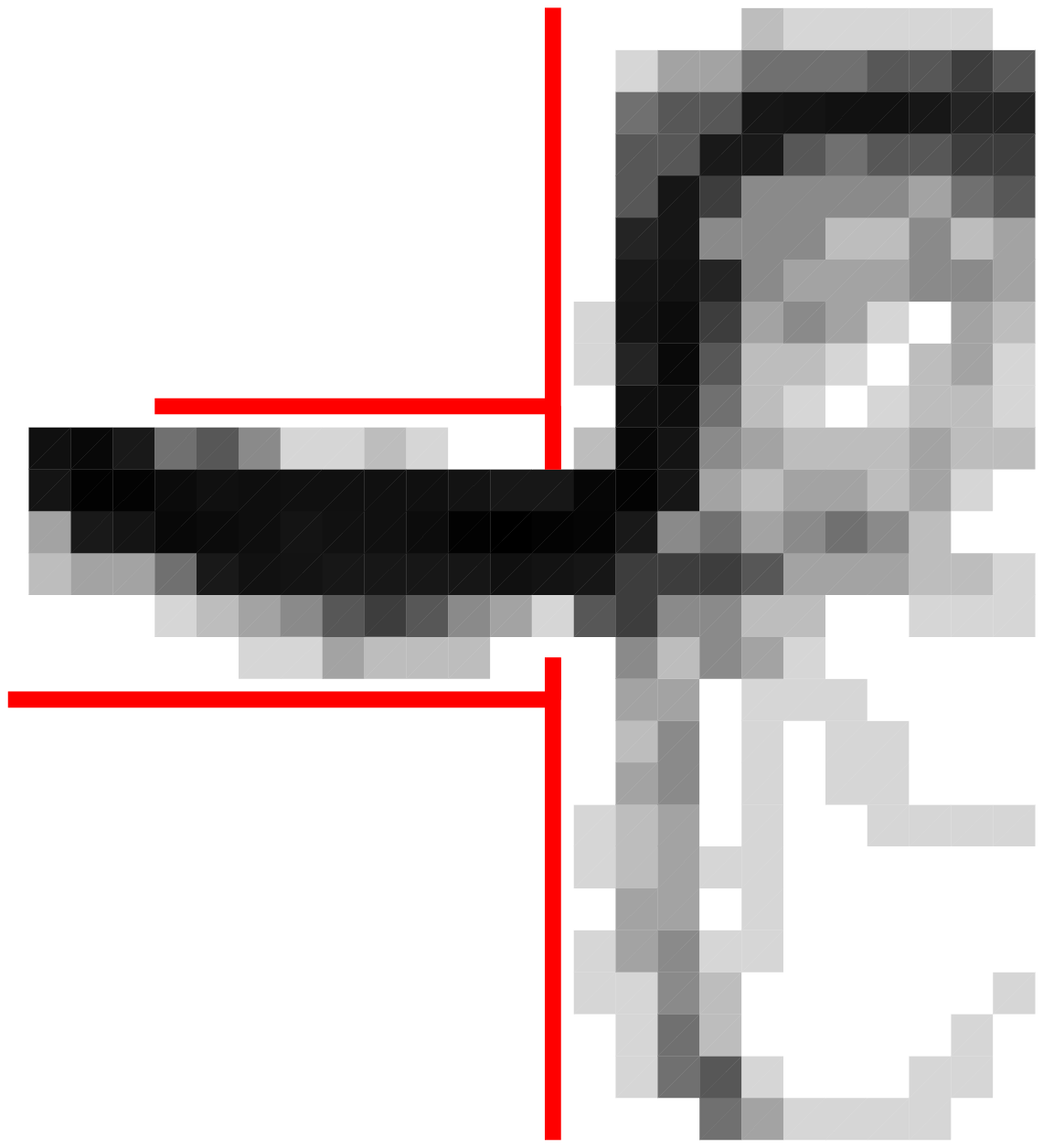}}&
	\framebox{\includegraphics[width=\iw, angle = 180, clip,trim=8cm 5cm 7cm 3cm]{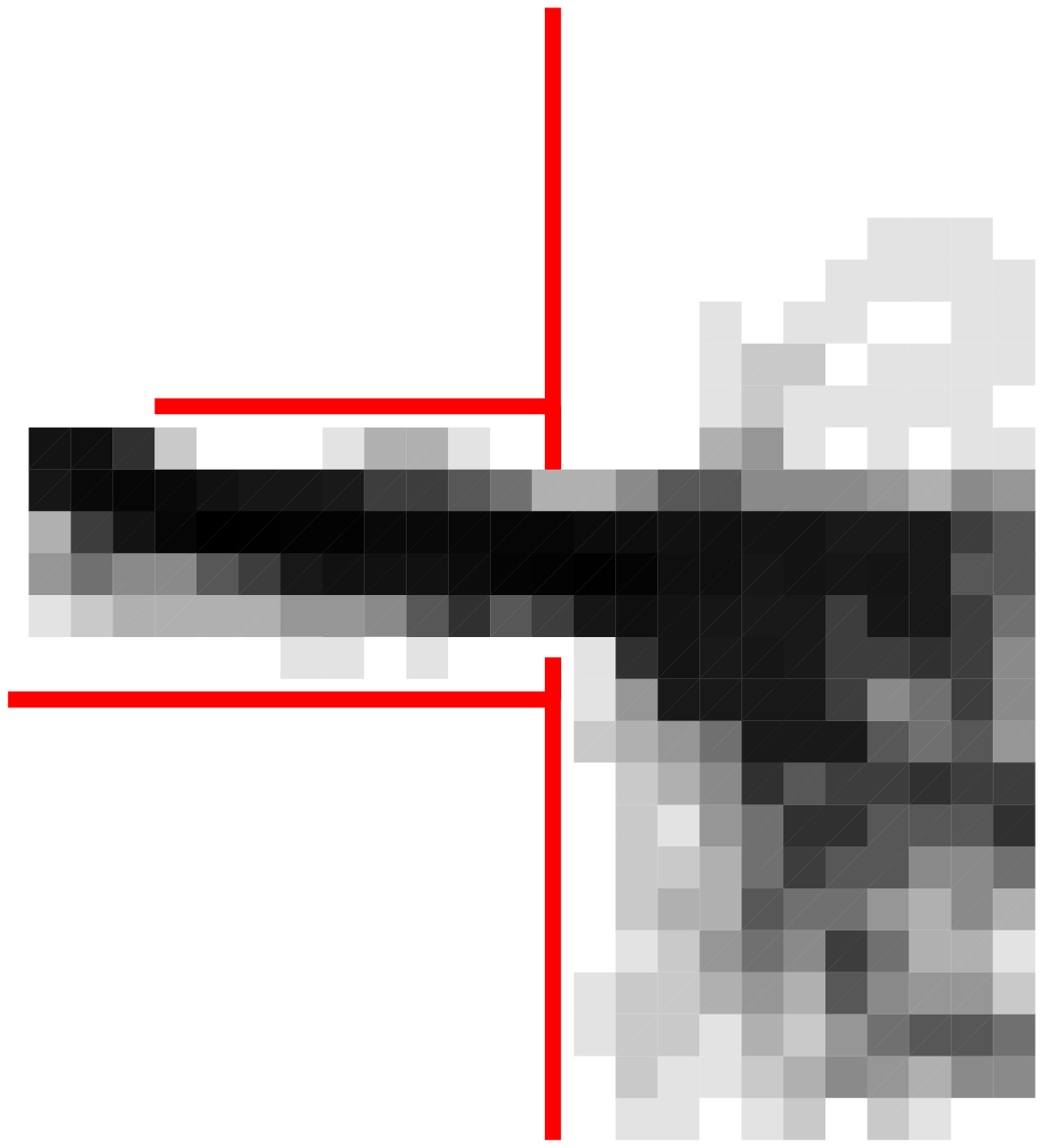}}
\end{tabular}
\caption{Paths (first row) and path density (second row) showing two conditions: metastable ($\overline{N}\in[25,35]$) and congested $\overline{N}\in[35,50]$. For both conditions, trajectories of slow and fast pedestrians are compared. The density was evaluated on a grid $0.2$~m $\times$ $0.2$~m. The darker the colour, the higher the density. Data from E4 were used.} 
\label{fig:PathDens}
\end{figure}

The route choice differences and patterns are visualised in Fig.~\ref{fig:PathDens}. The upper row shows all trajectories of given properties in one plot. To avoid optical illusion, we introduce the term path density, illustrating, how many trajectories pass through a given area.
More precisely, let the whole area $A$ of the room be artificially divided into disjoint sub-areas $A_j$ covering the room, i.e., $A=\bigcup_jA_j$. The path-density $\varrho(A_j)$~[path/m$^2$] of the sub-area $A_j$ is then defined as
\begin{equation}
	\varrho(A_j)=\frac{|W(A_j)|}{|A_j|}\,,\quad W(A_{j}) = \left\lbrace i\in I \mid \exists t\in T_i\,;\,(x_i(t),y_i(t)) \in A_{j} \right\rbrace\,,
\end{equation}
where $|\cdot|$ refers to the number of elements of a finite set or a size of a continuous set. The path density is given in the lower row of Fig.~\ref{fig:PathDens}.

For the purpose of discussing the issue, paths related to free-flow were excluded. The remaining paths were divided into four groups according to mean occupancy $\overline{N}$ and travel-time $TT$. Two conditions are distinguished: metastable for $\overline{N}\in[25,35]$ and congested $\overline{N}\in[35,50]$. The paths are then divided into slow and fast.

The graphs summarise and offer our observations. In the metastable regime, the crowd size enabled some participants to bypass the crowd, and thus their travel time was  significantly lower. However, in the congested regime, the crowd size affected the efficiency of the bypass strategy, and several fast participants pushed through the crowd directly in order to reach a low travel time. Further, we can see that fast participants preferred the right-hand side of the room, which can be due to asymmetry of the experiment outside the room or even imprinted behaviour.

The question is whether the slope of the linear model~(\ref{eq:LMTT}) is determined by the ability to push through the crowd, or rather by the willingness of participants to bypass the crowd. A deeper insight into this issue is provided by investigating the exit angle $\vartheta(i)$.

\section{Angle Analysis}

The exit angle $\vartheta(i)$ has been investigated mainly by using data from experiment E4. The conclusions were supported by data from the congested regime of experiment E2.

In this section two occupancy regimes are considered: without clogging in front of the exit (referred to as free-flow) and with clogging in front of the exit (referred to as congestion). Free-flow is characterized by the low occupancy $\overline{N}\leq15$, under which the interactions are present, but do not sufficiently affect the route-choice. Congestion is characterized by high occupancy $\overline{N}>15$, under which there is significant clogging in front of the exit, which motivates some participants to bypass the crowd. (Compare this division with that for the linear model, where the presence of interactions was important).

\begin{figure}[h!]
\begin{center}
\subfigure[Free flow]{
\resizebox*{6cm}{!}{\includegraphics[scale=1]{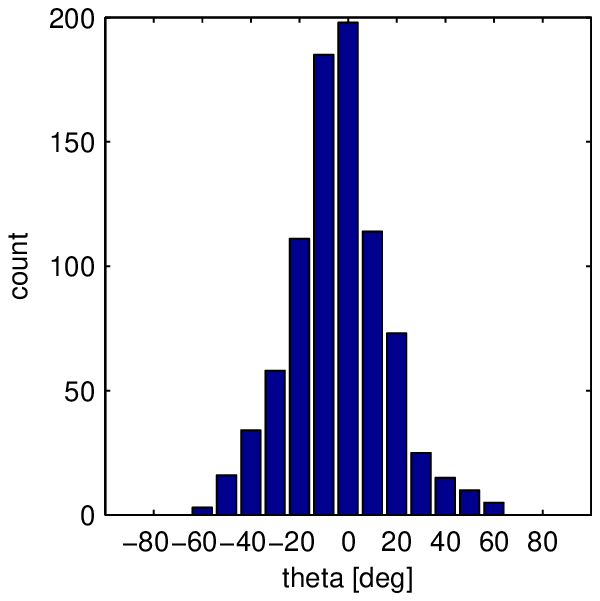}}}\hspace{5pt}
\subfigure[Congested]{
\resizebox*{6cm}{!}{\includegraphics[scale=1]{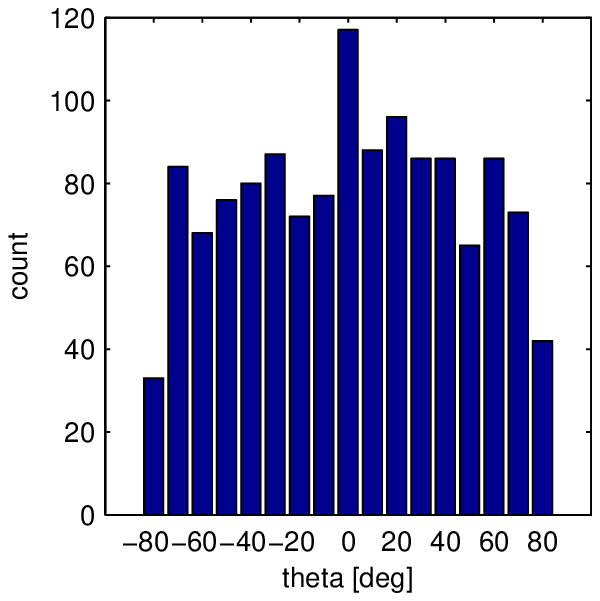}}}	\caption{Histograms of exit angles measured. Left: free flow $\overline{N}\leq15$. Right: congestion $\overline{N}>15$. Data from E4.}
\label{fig:histAngleE4}
\end{center}
\end{figure}

The frequency of path of given angles is depicted in Fig.~\ref{fig:histAngleE4}. As expected, in the free-flow regime exit angles $\vartheta(i)$ diverge from $0^{\circ}$ only rarely, $90$~\% of the angles are within the $(-45^{\circ},+45^{\circ})$ interval, which corresponds to the straight route between the entrance and the exit (note that there were three entrances, one centred, see Fig.~\ref{fig:ExpIlust}).

\begin{figure}[h!]
\begin{center}
\subfigure[FF]{
\resizebox*{6cm}{!}{\includegraphics[scale=1]{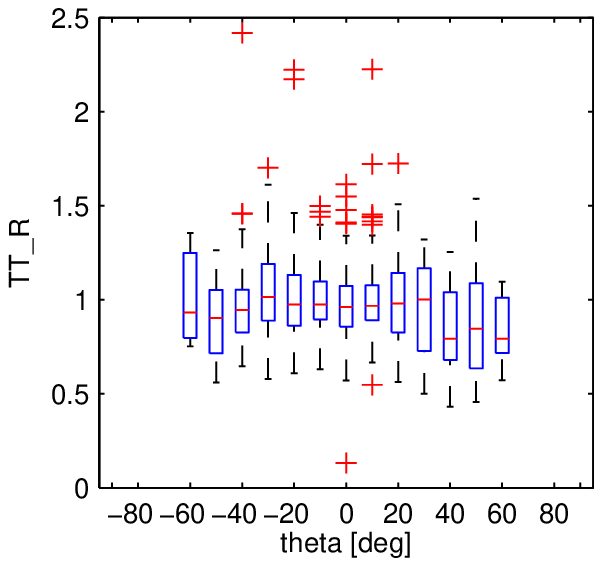}}}\hspace{5pt}
\subfigure[Congested]{
\resizebox*{6cm}{!}{\includegraphics[scale=1]{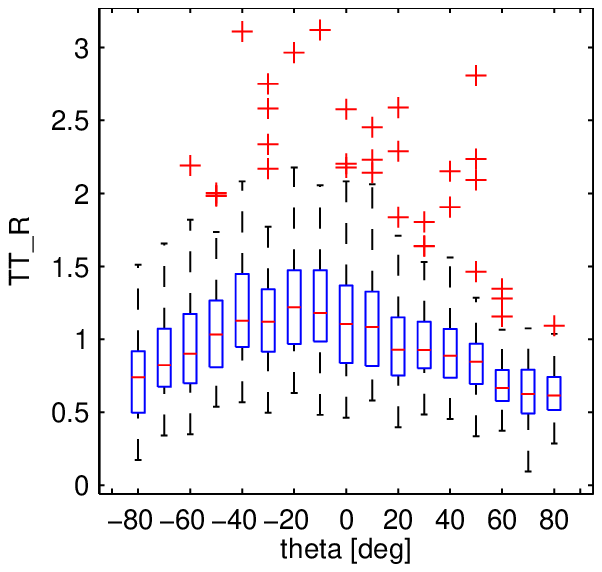}}}
\caption{Dependency of TT on exit angle sampled by 10 $\deg$. Data from E4.}
\label{fig:TTangleE4}
\end{center}
\end{figure}

In the congested regime, the angles from the whole range $(-90^{\circ},+90^{\circ})$ are distributed almost uniformly. Further information is provided in Figure~\ref{fig:TTangleE4}. Boxplots of relative travel-times are plotted against the exit angle. From these graphs we may conclude that on average the path approaching the exit under higher angle (in absolute value) reached a lower travel time. The observed asymmetry of the average travel-time may be related to the shape of the corridor behind the exit.

\begin{figure}[h!]
\begin{center}
\subfigure[Histogram]{
\resizebox*{6cm}{!}{\includegraphics[scale=1]{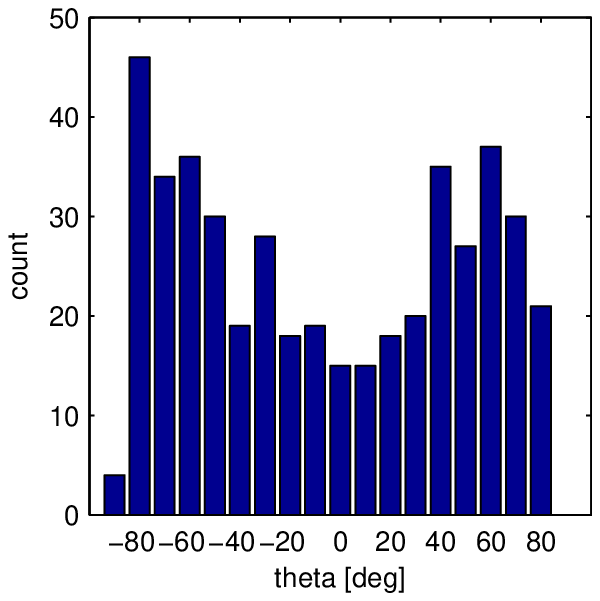}}}\hspace{5pt}
\subfigure[$TT$ -- exit angle dependency]{
\resizebox*{6cm}{!}{\includegraphics[scale=1]{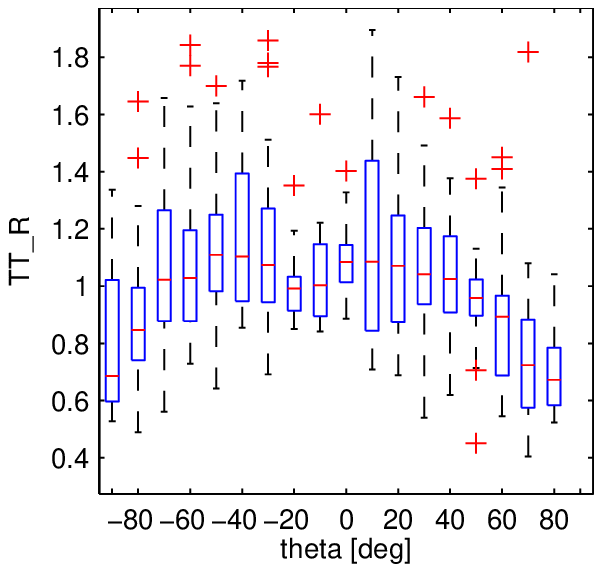}}}	\caption{Study of $TT$ with respect to exit angle, data from high density phase. Data from E2.}
\label{fig:angleE2}
\end{center}
\end{figure}

A very similar situation was observed during experiment E2, where the design was the same, but there was no corridor behind the exit. This modification may have caused no observable preferences of left- or right-hand side. In this experiment the paths with high $|\vartheta (i)|$ (related to motion along the walls) were even more frequent (see Figure \ref{fig:angleE2}). Here we note that the data of E2 stem from one run with rather high occupancy, and, therefore, all the paths were measured in the congestion regime.

Similar to to the travel-time analysis, we have also investigated the dependency of chosen angle $\vartheta$ and mean occupancy $\overline{N}$. At first, an individual linear model for each participant $\alpha$ has been  tested. We assume that for all paths $i\in I_\alpha$ the absolute value of $\vartheta(i)$ is given by linear model
\begin{equation}
	|\vartheta(i)|= \vartheta_{\alpha,0} + \overline{N}(i) \cdot \mathbf{c}_{\alpha} + \mathrm{noise}\,,
\end{equation}
where $\vartheta_{\alpha,0}$ can be interpreted as participant's preferred deviation from straight direction ($\vartheta_{\alpha,0}\geq0$), and $\mathbf{c}_{\alpha}$ is the slope of anticipated dependence.

\begin{figure}[h!]
\begin{center}
	{\includegraphics[scale=1]{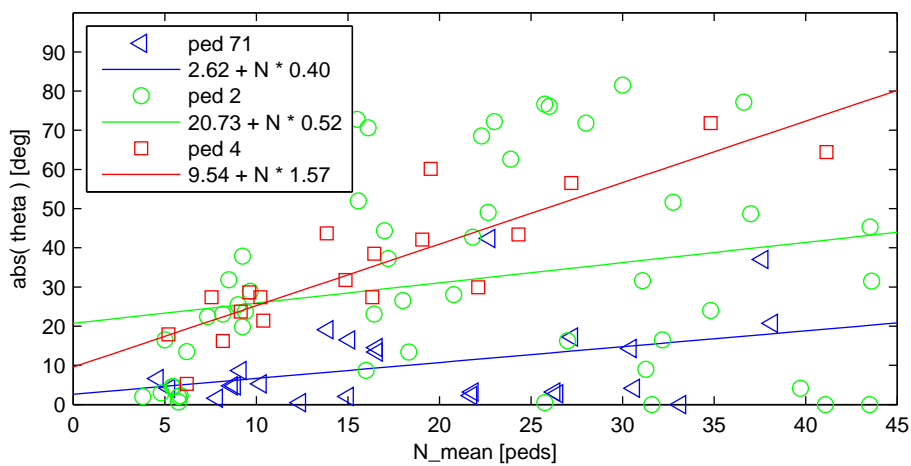}}
	\caption{Linear model $\vartheta$ -- $\bar{N}$ for selected representatives of different strategy groups.}
\label{fig:AngObs}
\end{center}
\end{figure}

Studying the $R^2$ statistics expressing, how much the use of the linear model decreases the variance in $\vartheta(i)$, we found that for majority of participants the statistics value is rather low ($R^{2}<0.3$). For visualization see Fig.~\ref{fig:AngObs}.

The linear model corresponds well to data related to participants with a consistent strategy, i.e. preferring bypassing the crowd (ped 4) or preferring direct pushing through the crowd (ped 71). Nevertheless, many participants tried to follow both strategies (ped 2). For those the linear model gives nonsensical results.

\section{Strategy Classification}

The previous section has shown that in contrast to $TT$ -- $\overline{N}$ dependence the linear model fails to characterize the dependence of the chosen angle on the mean occupancy. In this section we try to explain the relation between $\overline{N}(i)$, $\vartheta(i)$, and $TT(i)$ with respect to the chosen strategy. Our investigation is based on data from experiment E4.

\begin{figure}[h!]
\centering
	\includegraphics[scale=1]{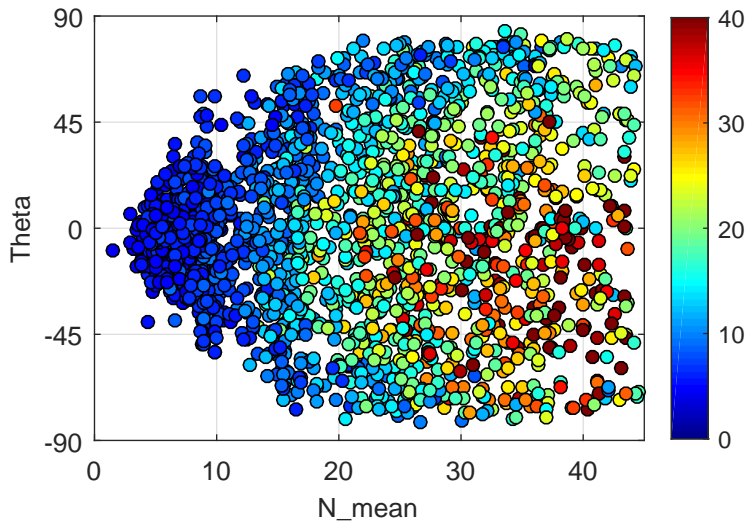}
\caption{Travel time visualized with respect to average occupancy and exit angle. Each point represents one path.}
\label{fig:AngNall}
\end{figure}

First, let us concentrate  information on all trajectories into one overview graph (see Fig.~\ref{fig:AngNall}). Each trajectory $i\in I$ is represented by one point; x-axis shows the average occupancy $\overline{N}(i)$, y-axis the exit angle $\vartheta(i)$, and the colour reflects the travel time $TT(i)$.

This supports the hypothesis that trajectories bypassing the crowd are ``faster'' than  those heading directly towards the exit. This effect is getting stronger if 
mean occupancy  $\overline{N}$ increases. Further, data clustering justifies the division into free-flow and congested regimes using the occupancy threshold $\overline{N}=15$.

Let us now have a closer look at individual participants. We assume that there are two possible route choices to be followed: either walking directly towards the exit regardless of the size of clogging or bypassing the crowd once clogging increases. In both cases , the pedestrian's travel time $TT$ is (1) or is not (0) significantly affected by the mean occupancy $\overline{N}$. Therefore, the pedestrians strategy can be described by a triplet
\begin{equation}
	\mathrm{Strategy}_\alpha=\big(\mathrm{direct}_\alpha,\mathrm{bypass}_{\alpha},\mathrm{preferred}_\alpha\big)\,,
\end{equation}
where $\mathrm{direct}$ and $\mathrm{bypss}$ are boolean variables denoting whether the participant's travel time is significantly affected by the occupancy, if walking directly or bypassing respectively, or not. The variable $\mathrm{preferred}\in\{\mathrm{direct},\mathrm{bypass},\mathrm{both}\}$ denotes the preferred route-choice of the participant.

The possible strategies can then be interpreted as follows:
\begin{itemize}
	\item (1,1,?) Slow in crowd regardless of route-choice.
	\item (1,0,?) Slow if pushing through directly, but fast if bypassing crowd.
	\item (0,0,?) Fast regardless of route-choice.
\end{itemize}
The frequency of strategies observed is given in Table~\ref{tab:class}. The classification has been performed by naked eye.

\begin{table}[h!]
\caption{Number of participants using strategies $\big(\mathrm{direct}_\alpha,\mathrm{bypass}_{\alpha},\mathrm{preferred}_\alpha\big)$. Data from E4.}
\label{tab:class}
\centering
	\begin{tabular}{clrl}
	\hline
	Type&Path&Count&\\
	\hline\hline
	$0,1$&--&0&\\
	\hline
	$0,0$&direct&$3$&\multirow{3}{*}{$\left.\begin{array}{c}\\ \\ \end{array}\right\}17$}\\
		 &bypass&$9$&\\
		 &both	&$5$&\\
	\hline
	$1,0$&direct&$6$&\multirow{3}{*}{$\left.\begin{array}{c}\\ \\ \end{array}\right\}19$}\\
		 &bypass&$9$&\\
		 &both	&$4$&\\
	\hline
	$1,1$&direct&$22$&\multirow{3}{*}{$\left.\begin{array}{c}\\ \\ \end{array}\right\}39$}\\
		 &bypass&$10$&\\
		 &both	&$7$&\\
	\hline
		
	\end{tabular}
\end{table}

\begin{figure}[h!]
	\includegraphics[scale=.8]{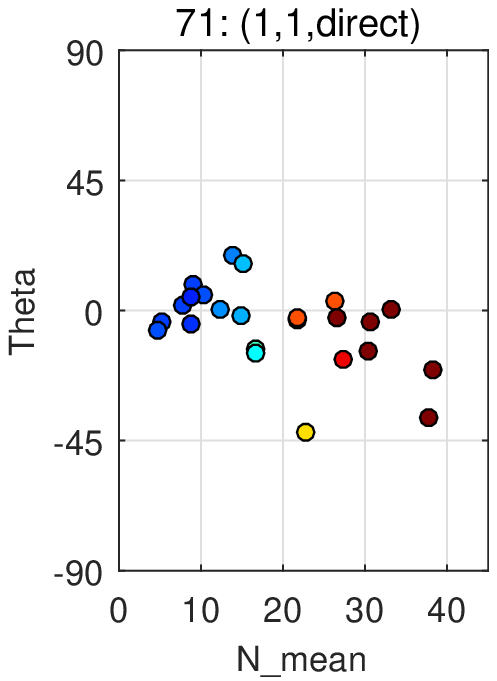}
	\hfill
	\includegraphics[scale=.8]{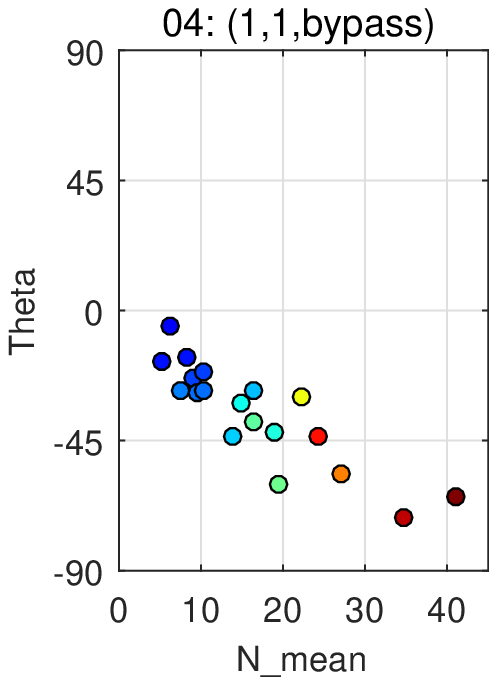} 
	\hfill
	\includegraphics[scale=.8]{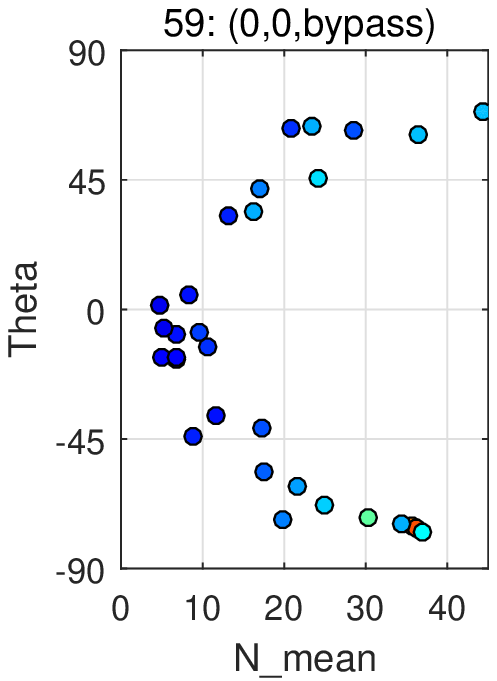}
\caption{$\vartheta-\overline{N}-TT$ graphs for three representatives of studied strategies. Colormap reflects $TT$ and is the same as in Fig.~\ref{fig:AngNall}}
\label{fig:repre1}
\end{figure}

\begin{figure}[h!]
	\includegraphics[scale=.8]{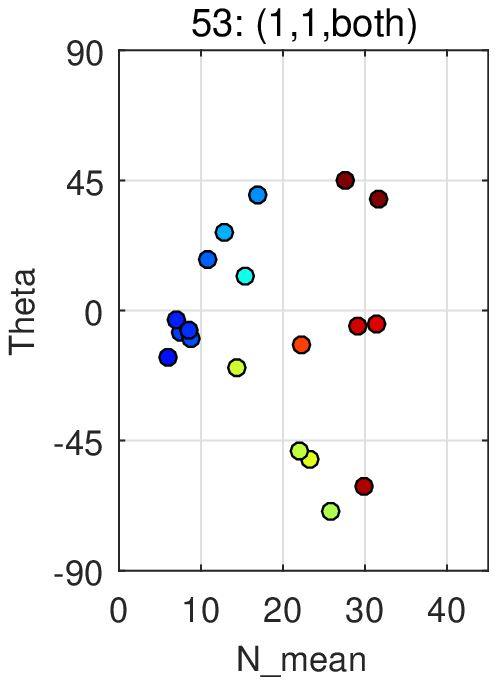}
	\hfill
	\includegraphics[scale=.8]{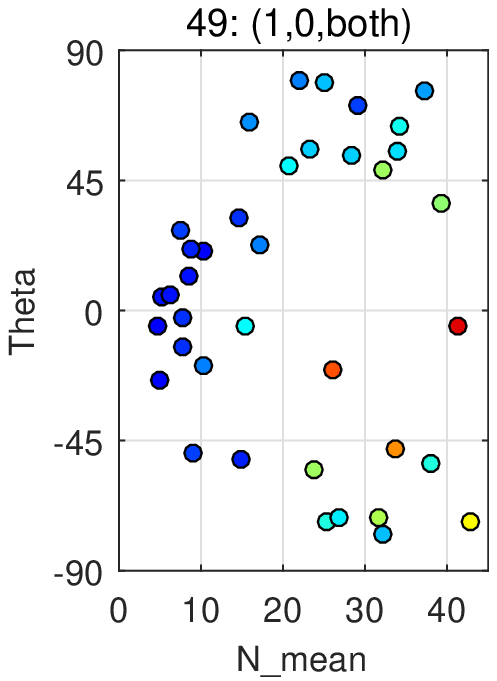}
	\hfill
	\includegraphics[scale=.8]{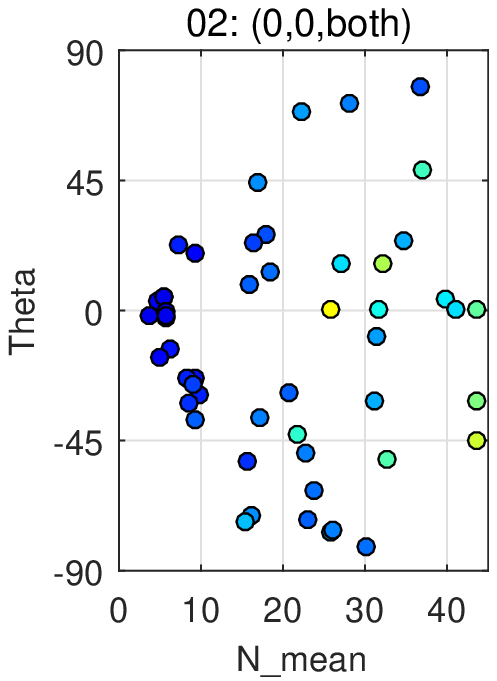} 
\caption{$\vartheta-\overline{N}-TT$ graphs for three representatives of combined strategies. Colormap reflects $TT$ and is the same as in Fig.~\ref{fig:AngNall}}
\label{fig:repre2}
\end{figure}

Three $\vartheta-\overline{N}-TT$ diagrams representing individual strategies are given in Fig. ~\ref{fig:repre1}. Three $\vartheta-\overline{N}-TT$ diagrams representing pedestrians with combined strategies are given in Fig.~\ref{fig:repre2}.

The above graphs and tables offer the following conclusions.
\begin{itemize}
	\item Approximately one half of the participants can be considered as non-aggressive in the sense that they neither push through the crowd nor try to bypass the crowd in order to reach a low travel time.
	\item Approximately one quarter of the participants can be considered as aggressive in the sense that they push directly through the crowd or bypass it effectively to reach low travel time.
	\item Approximately one quarter of the participants cannot push effectively through the crowd directly, but is successful in reaching a lower travel time if taking a bypass.
\end{itemize}

\section{Conclusions}

This article offers a deeper insight into the behaviour of pedestrians during egress situations through a narrow bottleneck. Due to automatic image recognition and multiple records of individual pedestrians, it was possible to analyse individual pedestrian’s behaviour under varying conditions.

Firstly, dependence of travel time on the number of pedestrians in the room has been modelled using a piece-wise linear model for each participant. The slope of linear dependence is closely associated with an individual’s ability to push through the crowd. This individual approach explains the high variance in relative travel time, which increases with increasing mean occupancy of the room.

This ability to push through the crowd is then further investigated by means of the angle under which the pedestrian joins the crowd in front of the exit. The study shows that the ability to push through the crowd can be partially associated with the aggressiveness of the individuals. 

Approximately one half of the pedestrians was not able to push through the crowd regardless of the chosen route. On the contrary, approximately one quarter was able to push effectively through the crowd directly or if bypassing the crowd. Such pedestrians can be denoted as aggressive in the sense that they are able to win conflicts with others. To support this idea, we found out many acts of aggressive behaviour from the camera records such as pushing, rude overtaking, or blocking each other. The concept of aggressiveness as pedestrian’s property fits this idea.

The remaining quarter was able to reach low travel time if bypassing the crowd, but was not able to push effectively through the crowd directly. Such pedestrians are not aggressive, but rather make the use the conditions for their benefit by reacting on the overcrowded area in front of the exit.

We consider this study to be applicable in microscopic simulations of pedestrian flow. Indeed, as shown in e.g.~\citep{HraBuk2017JCS}, introducing  heterogeneity into the ability to win conflicts and reaction into the occupancy enables to reproduce some patterns observable in pedestrian egress experiments, as e.g. line formation, bypassing the crowd, getting trapped in the clogging close to the exit door.

Using the boolean triplet to describe pedestrian strategy categorizes participants into 12 strategy groups. Generally, more smooth description and therefore less strict categorization could be defined, as e.g. continuous approach in all three elements. The coefficient of ``aggressiveness'' (slope of the linear model) quantifies the efficiency of the strategy chosen. A deeper analysis of this relation as well as the alternative strategy definition is a matter of further research.     

\section*{Acknowledgement}
This work was supported by the Czech Science Foundation under grant GA15-15049S and by the Czech Technical University under grant SGS15/214/ OHK4/3T/14.

Experimental records are available at https://www.youtube.com/watch?= d4zZpvhahYM.

All participants involved in the experiment gave their approval to processing and publishing the recorded materials for academic purposes.

\bibliographystyle{tfcad}
\bibliography{MBPH_ttra_ref}

\begin{thebibliography}{20}
\newcommand{\enquote}[1]{``#1''}
\providecommand{\natexlab}[1]{#1}
\providecommand{\url}[1]{\normalfont{#1}}
\providecommand{\urlprefix}{}

\bibitem[Bode and Codling(2016)]{BodCod2016TGF}
Bode, Nikolai W.~F., and Edward~A. Codling. 2016. \emph{Traffic and Granular
  Flow '15}, Chap. Statistical Models for Pedestrian Behaviour in Front of
  Bottlenecks, 81--88. Cham: Springer International Publishing.

\bibitem[Buk{\'a}{\v{c}}ek, Hrab{\'a}k, and
  Krb{\'a}lek(2015)]{BukHraKrb2015TGF}
Buk{\'a}{\v{c}}ek, Marek, Pavel Hrab{\'a}k, and Milan Krb{\'a}lek. 2015.
  \emph{Traffic and Granular Flow '13}, Chap. Experimental Analysis of
  Two-Dimensional Pedestrian Flow in Front of the Bottleneck, 93--101. Cham:
  Springer International Publishing.

\bibitem[Buk{\'a}{\v{c}}ek, Hrab{\'a}k, and
  Krb{\'a}lek(2016)]{BukHraKrb2016TGF}
Buk{\'a}{\v{c}}ek, Marek, Pavel Hrab{\'a}k, and Milan Krb{\'a}lek. 2016.
  \emph{Traffic and Granular Flow '15}, Chap. Individual Microscopic Results Of
  Bottleneck Experiments, 105--112. Cham: Springer International Publishing.

\bibitem[Buk\'a\v{c}ek, Hrab\'ak, and Krb\'alek(2014)]{BukHraKrb2014Procedia}
Buk\'a\v{c}ek, Marek, Pavel Hrab\'ak, and Milan Krb\'alek. 2014.
  ``{Experimental Study of Phase Transition in Pedestrian Flow}.'' In
  \emph{Pedestrian and Evacuation Dynamics 2014},  edited by W.~Daamen, D.~C.
  Duives, and S.~P. Hoogendoorn, Vol.~2 of \emph{Transportation Research
  Procedia}, 105--113. Elsevier Science B.V.

\bibitem[Campanella, Hoogendoorn, and Daamen(2009)]{CamHooDaa2009TRR}
Campanella, M., S.P. Hoogendoorn, and W.~Daamen. 2009. ``Effects of
  heterogeneity on self-organized pedestrian flows.'' \emph{Transportation
  Research Record}  (2124): 148--156.

\bibitem[Duives, Daamen, and Hoogendoorn(2014)]{DuiDaaHoo2014Procedia}
Duives, Dorine, Winnie Daamen, and Serge Hoogendoorn. 2014. ``{Anticipation
  Behavior Upstream of a Bottleneck}.'' In \emph{Pedestrian and Evacuation
  Dynamics 2014},  edited by W.~Daamen, D.~C. Duives, and S.~P. Hoogendoorn,
  Vol.~2 of \emph{Transportation Research Procedia}, 43--50. Elsevier Science
  B.V.

\bibitem[Haghani and Sarvi(2016)]{HagSar2016TRR}
Haghani, Milad, and Majid Sarvi. 2016. ``Identifying Latent Classes of
  Pedestrian Crowd Evacuees.'' \emph{Transportation Research Record: Journal of
  the Transportation Research Board} 2560: 67--74.

\bibitem[Helbing et~al.(2005)]{HelBuzJohWer2005TS}
Helbing, Dirk, Lubos Buzna, Anders Johansson, and Torsten Werner. 2005.
  ``Self-Organized Pedestrian Crowd Dynamics: Experiments, Simulations, and
  Design Solutions.'' \emph{Transportation Science} 39 (1): 1--24.

\bibitem[Hrab{\' a}k and Buk{\' a}\v{c}ek(2017)]{HraBuk2017JCS}
Hrab{\' a}k, P., and M.~Buk{\' a}\v{c}ek. 2017. ``{Influence of agents
  heterogeneity in cellular model of evacuation}.'' \emph{Journal of
  Computational Science} 21: 486--493.

\bibitem[Kretz, Gr\"unebohm, and Schreckenberg(2006)]{KreGruSch2006JSM}
Kretz, Tobias, Anna Gr\"unebohm, and Michael Schreckenberg. 2006.
  ``Experimental study of pedestrian flow through a bottleneck.'' \emph{Journal
  of Statistical Mechanics: Theory and Experiment} 2006 (10): P10014/1--20.

\bibitem[Liao et~al.(2016)]{LiaTorSeyChrDrzZheZha2016PhysicaA}
Liao, Weichen, Antoine Tordeux, Armin Seyfried, Mohcine Chraibi, Kevin
  Drzycimski, Xiaoping Zheng, and Ying Zhao. 2016. ``Measuring the steady state
  of pedestrian flow in bottleneck experiments.'' \emph{Physica A: Statistical
  Mechanics and its Applications} 461: 248--261.

\bibitem[Mehner et~al.(2015)]{MehBolMatLei2015LNCS}
Mehner, W., M.~Boltes, M.~Mathias, and B.~Leibe. 2015. ``{Robust marker-based
  tracking for measuring crowd dynamics}.'' In \emph{Computer Vision Systems:
  10th International Conference, ICVS 2015, Copenhagen, Denmark, July 6-9,
  2015, Proceedings},  edited by Lazaros Nalpantidis, Volker Kr{\"u}ger,
  Jan-Olof Eklundh, and Antonios Gasteratos, Vol. 9163 of \emph{Lecture Notes
  in Computer Science}, 445--455. Cham: Springer International Publishing.

\bibitem[Mehner, Boltes, and Seyfried(2016)]{MehBolSey2016TGF}
Mehner, W., M.~Boltes, and A.~Seyfried. 2016. \emph{Traffic and Granular Flow
  '15}, Chap. Methodology for Generating Individualised Trajectories from
  Experiments, 3--10. Cham: Springer International Publishing.

\bibitem[Parisi and Dorso(2005)]{ParDor2005PhysA}
Parisi, D.R., and C.O. Dorso. 2005. ``Microscopic dynamics of pedestrian
  evacuation.'' \emph{Physica A: Statistical Mechanics and its Applications}
  354: 606--618.

\bibitem[Schadschneider, Chowdhury, and Nishinari(2010)]{SchChoNis2010}
Schadschneider, A., D.~Chowdhury, and K.~Nishinari. 2010. \emph{Stochastic
  Transport in Complex Systems: From Molecules to Vehicles}. Amsterdam:
  Elsevier Science B. V.

\bibitem[Schadschneider et~al.(2009)]{SchKliKluKreRogSey2009}
Schadschneider, Andreas, Wolfram Klingsch, Hubert Kl{\"u}pfel, Tobias Kretz,
  Christian Rogsch, and Armin Seyfried. 2009. \emph{Encyclopedia of Complexity
  and Systems Science}, Chap. {Evacuation Dynamics: Empirical Results, Modeling
  and Applications}, 3142--3176. New York, NY: Springer New York.

\bibitem[Seyfried et~al.(2009)]{SeyPasSteBolRupKli2009TS}
Seyfried, Armin, Oliver Passon, Bernhard Steffen, Maik Boltes, Tobias
  Rupprecht, and Wolfram Klingsch. 2009. ``{New Insights into Pedestrian Flow
  Through Bottlenecks}.'' \emph{Transportation Science} 43 (3): 395--406.

\bibitem[Seyfried, Portz, and Schadschneider(2010)]{SeyPorSch2010LNCS}
Seyfried, Armin, Andrea Portz, and Andreas Schadschneider. 2010. ``Phase
  Coexistence in Congested States of Pedestrian Dynamics.'' In \emph{Cellular
  Automata},  edited by Stefania Bandini, Sara Manzoni, Hiroshi Umeo, and
  Giuseppe Vizzari, Vol. 6350 of \emph{Lecture Notes in Computer Science},
  496--505. Springer Berlin Heidelberg.

\bibitem[Steffen and Seyfried(2010)]{SteSey2010PhysA}
Steffen, B., and A.~Seyfried. 2010. ``Methods for measuring pedestrian density,
  flow, speed and direction with minimal scatter.'' \emph{Physica A:
  Statistical Mechanics and its Applications} 389 (9): 1902--1910.

\bibitem[Zhang and Seyfried(2014)]{ZhaSey2014Procedia}
Zhang, Jun, and Armin Seyfried. 2014. ``{Quantification of Bottleneck Effects
  for Different Types of Facilities}.'' In \emph{Pedestrian and Evacuation
  Dynamics 2014},  edited by W.~Daamen, D.~C. Duives, and S.~P. Hoogendoorn,
  Vol.~2 of \emph{Transportation Research Procedia}, 51--59. Elsevier Science
  B.V.

\end{thebibliography}

\end{document}